\documentclass[twocolumn]{aastex63}

\usepackage{txfonts}

\usepackage{color,bm}

\newcommand{\eqref}[1]{(\ref{#1})}

\shorttitle{Super-Rayleigh Slopes Generated by Photochemical Haze}
\shortauthors{Ohno \& Kawashima}

\begin{document}

%\title{Emergence of Super-Rayleigh Slopes in Transmission Spectra of Exoplanets: Insight from Haze Vertical Profiles}

\title{Super-Rayleigh Slopes in Transmission Spectra of Exoplanets Generated by Photochemical Haze}

%\title{Haze-driven Steep Slope Formation in Transmission Spectra of Exoplanets}

\author[0000-0003-3290-6758]{Kazumasa Ohno}
\affil{Department of Earth and Planetary Sciences, Tokyo Institute of Technology, Meguro, Tokyo, 152-8551, Japan}

\author[0000-0003-3800-7518]{Yui Kawashima}
\affil{SRON Netherlands Institute for Space Research, Sorbonnelaan 2, 3584 CA Utrecht, The Netherlands}
%\affil{Department of Earth and Planetary Sciences, Tokyo Institute of Technology, Meguro, Tokyo, 152-8551, Japan}

\begin{abstract}
%Transmission spectra of exoplanetary atmospheres sometimes show optical spectral slopes that are too steep to be explained by the canonical Rayleigh scattering slope.
Spectral slopes in {optical} transmission spectra of exoplanetary atmospheres encapsulate information on the properties of exotic clouds.
The slope is usually attributed to the Rayleigh scattering caused by tiny {aerosol} particles, whereas recent retrieval studies have suggested that the slopes are often steeper than the canonical Rayleigh slopes.
Here, we propose that photochemical haze formed in vigorously mixing atmospheres can explain such super-Rayleigh slopes.
We first analytically show that the spectral slope can be steepened by the vertical opacity gradient in which atmospheric opacity increases with altitude.
%produces the slope quite steeper than the Rayleigh slope. 
Using a microphysical model, we demonstrate that such opacity gradient can be naturally generated by photochemical haze, especially when the eddy mixing is substantially efficient.
The transmission spectra of hazy atmospheres can be demarcated into four typical regimes in terms of the haze mass flux and eddy diffusion coefficient.
We find that the transmission spectrum can have the spectral slope $2$--$4$ times steeper than the Rayleigh slope if the eddy diffusion coefficient is sufficiently high and the haze mass flux falls into a moderate value.
%{Based on current theoretical understanding on the eddy diffusion, we suggest that photochemical haze preferentially generates super-Rayleigh slopes at planets with equilibrium temperature of $1000$--$1500~{\rm K}$, which might be consistent with results of recent retrieval studies.}
Based on the eddy diffusion coefficient suggested by a recent study of atmospheric circulations, we suggest that photochemical haze preferentially generates super-Rayleigh slopes at planets with equilibrium temperature of $1000$--$1500~{\rm K}$, which might be consistent with results of recent retrieval studies.
Our results would help to interpret the observations of spectral slopes from the perspective of haze formation.
\end{abstract}

%\keywords{planets and satellites: atmospheres -- planets and satellites: composition -- planets and satellites: individual(Kepler-51b, Kepler-51c)}

%%%%%%%%%%%%%%%%%%%%%
\section{Introduction} \label{sec:intro}
Transmission spectroscopy is a powerful way to explore the properties of exoplanetary atmospheres \citep[e.g.,][]{Charbonneau+02,Pont+13,Kreidberg+14,Sing+16}.
%A number of studies have observed the transmission spectra and constrained the atmospheric composition as well as properties of exotic clouds \citep[e.g.,][]{Charbonneau+02,Pont+13,Kreidberg+14,Sing+16}.
One of the remarkable feature{s} of the transmission spectra is the rise of transit depth toward blue in {the optical wavelength}, called a spectral slope.
The slope is quantified by \citep[e.g.,][]{Levavelier+08}
\begin{equation}\label{eq:slope0}
   % \frac{dD}{d\log{\lambda}}\propto H\alpha,
   \frac{dR_{\rm p}}{d\ln{\lambda}}= H\alpha,
\end{equation}
where $R_{\rm p}$ is the planetary radius, $\lambda$ is the wavelength, $H$ is the pressure scale height, and $\alpha$ is the spectral index of atmospheric opacity, i.e., $\kappa\propto \lambda^{\alpha}$.
Utilizing Equation \eqref{eq:slope0}, measurements of the slopes can help to constrain atmospheric properties as well as the properties of exoplanetary clouds.
%The slope is often attributed to the Rayleigh scattering caused by tiny condensate particles \citep[e.g.,][]{Levavelier+08}.

%Recent retrieval studies suggested that the slope is quite steeper than the theoretical prediction in many cases.
%Although the slope measurement has been extensively conducted, its interpretation is not straightforward.
%In particular, many studies assumed $\alpha=-4$ assuming the Rayleigh scattering of tiny aerosol particles.
%For example, $\alpha$ takes $\alpha=-4$ for Rayleigh scattering of tiny particles, $\alpha=0$ for gray large particles, and $\alpha=-2$ for scattering of porous aggregates \citep[e.g.,][]{Ohno+19}.
%The spectral slopes are often attributed to the Rayleigh scattering caused by tiny aerosol particles, and the spectral index of $\alpha=-4$ is assumed \citep[e.g.,][]{Barstow+17}.
The spectral slopes are often attributed to the Rayleigh scattering ($\alpha=-4$) caused by tiny aerosols; however, several exoplanets actually exhibit the slopes steeper than the Rayleigh slope.
%However, several exoplanets actually exhibit the slopes quite steeper than the Rayleigh slope.
A retrieval studies of \citet{Pinhas+19} { and \citet{Welbanks+19}} obtained the median spectral index of $\alpha\la-5$ for most of the transmission spectra of hot Jupiters collected by \citet{Sing+16}.
%Even steeper spectral slopes, say $\alpha<-20$, have been reported for several hot exoplanets \citep{Sedaghati+17,May+20}.
For a more extreme example, \citet{Sedaghati+17} showed that the slope of a hot Jupiter WASP-19b is characterized by $\alpha=-26$.
\citet{May+20} found an even steeper spectral slope { of $\alpha\approx-35$} for a hot Neptune HATS-8b.

Several mechanisms potentially explain such ``super-Rayleigh slopes'' (SRSs hereafter).
For example, unocculted star spots produce steep slope-like feature in transmission spectra \citep[e.g.,][]{McCullough+14} and potentially explain some SRSs.
In fact, \citet{Espinoza+19} reported that the effects of star spots can explain the SRS of WASP-19b observed by \citet{Sedaghati+17}.
%However, unocculted star spots would not always explain the nature of SRS, especially for a planet around an inactive star, as argued in \citep{May+20}.
%Alternatively, NUV absorving gasses, such as SH, can yield steep slope-like feature in the {wavelength shortward of $\sim0.46~{\rm \mu m}$} \citep{Zahnle+09,Evans+18}.
The mercapto radical, SH, can also yield steep slope-like feature, though it is responsible only for near the NUV wavelength \citep[$\lambda\la0.46~{\rm \mu m}$,][]{Zahnle+09,Evans+18}.
Alternatively, tiny sulphide condensates, such as MnS, can produce slope-like feature with $\alpha < -5$ \citep{Pinhas&Madhusudhan17}.
Most recently, \citet{Kawashima&Ikoma19} found that photochemical haze can steepen the Rayleigh slope if atmospheric eddy diffusion is efficient.
%{Efficient eddy diffusion can steepen the Rayleigh scattering slope in the atmospheres with photochemical haze---aerosols formed via photochemistry in upper atmospheres \citep{Kawashima&Ikoma19}.}

%In this letter, we propose an alternative idea explaining SRSs: vertical opacity gradient generated by photochemical haze---aerosols formed via photochemistry in upper atmospheres---can produce the slope quite steeper than the canonical Rayleigh slope.
{In this study, we generalize the conditions in which photochemical haze produces the steep spectral slopes.}
{In Section \ref{sec:method}, we analytically show that the vertical opacity gradient can steepen the spectral slope than the canonical Rayleigh slope. We also demonstrate that photochemical haze can generate such opacity gradient.}% if the eddy diffusion coefficient is sufficiently high.}
In Section \ref{sec:result}, we calculate the synthetic transmission spectra of hazy atmospheres for a wide range of eddy diffusion coefficient and haze mass flux, and discuss {the conditions of these parameters for} {which the SRSs emerge}.
%We also show that the transmission spectra of hazy atmospheres can be demarcated into four typical regimes.
In Section \ref{sec:summary}, we summarize our findings.

%the assumption could be violated for aerosol opacity if atmospheric mixing is significantly strong.

%The organization of this paper is as follows.
%In Section \ref{sec:method}, we 

%%%%%%%%%%%%%%%%%%%%%
\section{A mechanism producing steep spectral slopes by haze} \label{sec:method}
%\subsection{Vertical opacity gradient steepens the spectral slope}\label{eq:derivation_slope}
%We first investigate how the spectral slope depends on the vertical gradient of atmospheric opacity.
{A key factor producing the SRSs is the vertical gradient of atmospheric opacity.}
{This fact is not captured} by Equation \eqref{eq:slope0} that was derived under the assumption of vertically uniform opacity. 
%, whereas the assumption is not necessary valid.
%The optical depth of a radial path length $dr$ is given by
%\begin{equation}\label{eq:dtau}
%    d\tau=\rho_{\rm g}(r)\kappa dr,
%\end{equation}
%where $\rho_{\rm g}$ is the atmospheric density. 
To examine the effect of vertical opacity gradient, we {assume the opacity following} $\kappa=\kappa_{\rm 0}(\lambda/\lambda_{\rm 0})^{\alpha}(P/P_{\rm 0})^{-\beta}$, where $P$ is the atmospheric pressure
{and $\kappa_{\rm 0}$ is the opacity at the pressure level of $P_{\rm 0}$ { and the wavelength of $\lambda_{\rm 0}$}.}
%, in which the opacity is higher at higher altitude for $\beta>0$ and vice versa for $\beta<0$.
{Assuming hydrostatic equilibrium, and constant temperature and gravity throughout the atmosphere,} the chord optical depth {at the impact paramter of $r$} is calculated as \citep[e.g.,][]{2012ApJ...753..100B}
\begin{equation}\label{eq:tauv_0}
    \tau_{\rm v}(r)={2} \rho_{\rm g}(r)\kappa(r)\int_{{r}}^{\infty}\exp{\left[ -(1-\beta)\frac{r'-r}{H}\right]}\frac{r'dr'}{\sqrt{r'^2-r^2}},
\end{equation}
%{\begin{equation}\label{eq:tauv_0}
%    \tau_{\rm v}(r)={2} \rho_{\rm g}(r)\kappa(r)\int_{{0}}^{\infty}\exp{\left[ -(1-\beta)\frac{r'-r}{H}\right]}dx,
%\end{equation}}
where $\rho_{\rm g}$ is the atmospheric density, and $r'$ is the radial distance from the center of the planet.
Applying the transformation of {$x=\sqrt{r^2-r'^2}$} and {approximation of} $r'-r\approx x^2/2r$ as in \citet{Fortney05}, Equation \eqref{eq:tauv_0} is rewritten as
\begin{equation}\label{eq:tauv_int}
    \tau_{\rm v}(r)=\rho_{\rm g}(r)\kappa(r)\int_{\rm -\infty}^{\rm \infty}\exp{\left[ -\frac{(1-\beta)x^2}{2rH}\right]}dx.
\end{equation}
Equation \eqref{eq:tauv_int} diverges for $\beta\geq1$, and {thus} a finite region of the atmosphere, in which the opacity source exists, {should be taken} into account for $\beta\geq1$.
Here, we see the solution of Equation \eqref{eq:tauv_int} only for $\beta<1$. %to keep a simplicity for subsequent argument. %\footnote{If we solve Equation \eqref{eq:tauv_int} for the finite atmospheric region, the chord optical depth is expressed by the error function ($\beta<1$) or the imaginary error function ($\beta>1$). }.
As shown later, the opacity gradient produced by photochemical haze is mostly characterized by $\beta<1$.
%yk{since the particles settle down faster in the lower atmosphere due to their larger sizes in the lower atmosphere}.
For $\beta<1$, the chord optical depth is calculated as
\begin{equation}\label{eq:tauv}
    \tau_{\rm v}(r) {=} \rho_{\rm 0}\kappa_{\rm 0}\left(\frac{\lambda}{\lambda_{\rm 0}} \right)^{\alpha}\sqrt{\frac{2{\pi}R_{\rm 0}H}{1-\beta}}\exp{\left[ -(1-\beta)\frac{r-R_{\rm 0}}{H}\right]},
\end{equation}
where $\rho_{\rm 0}$ is the atmospheric density at the reference radius of $R_{\rm 0}$.
The observed planetary radius is corresponding to the radius at $\tau_{\rm v}\sim1$.
%\footnote{The threshold optical depth is actually not exactly unity \citep{Levavelier+08,Heng&Kitzmann17}. However, the choice of the threshold optical depth does not affect the gradient of the spectral slope.}.
{Inserting $\tau_{\rm v}=1$ in} Equation \eqref{eq:tauv}, the observed radius is given by
\begin{equation}\label{eq:z_h}
    R_{\rm p}=R_{\rm 0}+\frac{H}{1-\beta}\ln{\left[ \rho_{\rm 0}\kappa_{\rm 0}\left(\frac{\lambda}{\lambda_{\rm 0}} \right)^{\alpha}\sqrt{\frac{2\pi R_{\rm 0}H}{1-\beta}}\right]}.
\end{equation}
Differentiating Equation \eqref{eq:z_h} with respect to $\lambda$, we achieve a spectral slope with vertical opacity gradient applicable for $\beta<1$:
\begin{equation}\label{eq:slope}
    \frac{dR_{\rm p}}{d\ln{\lambda}}= \frac{H\alpha}{1-\beta}.
\end{equation}
Equation \eqref{eq:slope} is essentially the same as Equation \eqref{eq:slope0} except for the factor of $(1-\beta)^{-1}$.

%If we assume that the opacity source exists only at $P>P_{\rm top}$, the integration is carried out to $x= \sqrt{2Hr\ln{(P/P_{\rm top})}}$.
%Then, the chord optical depth is calculated as
%\begin{equation}\label{eq:tauv_fin}
%    \tau_{\rm v}(r) = \rho_{\rm g}(r)\kappa(r)\sqrt{2\pi rH}\times
%\left\{
%\begin{array}{ll}
%      {\displaystyle {\frac{ {\rm erf}\left(\sqrt{(1-\beta)\ln{\left(P/P_{\rm top}\right)}} \right) }{\sqrt{1-\beta}}}  & \text{for $\beta < 1$} }\\[3ex]
%      {\displaystyle \sqrt{(4/\pi)\ln{(P/P_{\rm top})}} & \text{for $\beta = 1$} }\\[1.5ex]
%     {\displaystyle {\frac{ {\rm erfi}\left(\sqrt{(\beta-1)\ln{\left(P/P_{\rm top}\right)}} \right) }{\sqrt{\beta-1}}}  & \text{for $\beta > 1$} }
%         \end{array}
%\right.,
%\end{equation}
%where $\rm erf(x)$ is the error function, and $\rm erfi(x)$ is the imaginary error function.
%Equation \eqref{eq:tauv_fin} restores a well-used formula of $\tau_{\rm v}=\rho_{\rm g}\kappa \sqrt{2\pi rH}$ for $\beta=0$ and $P_{\rm top}=0$.

%The most important implication of Equation \eqref{eq:slope} is that 
An important implication of Equation \eqref{eq:slope} is that the spectral index $\alpha$ (or scale height $H$) is degenerated with vertical opacity gradient $\beta$. 
Notably, for ${0<\beta<1}$ in which the opacity is higher at higher altitude, the slope is steepened by a factor of $(1-\beta)^{-1}$ from the classical prediction of Equation \eqref{eq:slope0}.
Thus, it is crucial to take into account the vertical opacity gradient to explore the nature of SRSs.
%The remaining question is what causes the circumstance of $\beta>0$.
%In next subsection, we show that photochemical hazes may yield the vertical opacity gradient of $\beta>0$.

%\subsection{Opacity Gradient generated by Photochemical Hazes}\label{eq:derivation_haze}
The remaining question is what causes the opacity gradient {with $0<\beta<1$}.
{ We suggest that photochemical haze {can} naturally produce such gradient. 
As shown in Appendix \ref{appendix1}, for haze particles much smaller than the gas mean free path and the relevant wavelength, the opacity can be written as
{\begin{equation}\label{eq:kappa}
    \kappa =\frac{36\pi gHF}{\rho_{\rm p}} {\underbrace{\frac{1}{Pv_{\rm t}}\left[1-\exp{\left(-\frac{v_{\rm t}H}{K_{\rm z}} \right)}\right]}_{\text{Pressure dependence}}}\underbrace{\frac{nk\lambda^{-1}}{(n^2-k^2+2)^2+(2nk)^2}}_{\text{Wavelength dependence}},
\end{equation}}
where $F$ is the haze mass flux, $K_{\rm z}$ is the eddy diffusion coefficient, $g$ is the surface gravity, $\rho_{\rm p}$ is the particle density, and $n$ and $k$ are the real and imaginary parts of complex refractive index. 
$v_{\rm t}$ is the terminal velocity of haze particles approximated by \citep{Woitke&Helling03}
\begin{equation}\label{eq:vt_epstein}
    v_{\rm t}\approx \frac{\rho_{\rm p}g^2H}{P\sqrt{8k_{\rm B}T/\pi m_{\rm g}}}a,
    \label{eq:vt}
\end{equation}
where $k_{\rm B}$ is the Boltzmann constant, $T$ is the temperature, $m_{\rm g}$ is the mean mass of atmospheric gas particles, and $a$ is the particle radius.
The asymptotic behaviors of Equation \eqref{eq:kappa} clarify the pressure dependence as
\begin{equation}\label{eq:asympt}
    \kappa \propto 
\left\{
\begin{array}{lr}
      P^{-1}  & \text{($v_{\rm t}H/K_{\rm z} \ll 1$)} \\
      a^{-1}\rho_{\rm p}^{-1} & \text{($v_{\rm t}H/K_{\rm z} \gg 1$)}.
         \end{array}
\right.
\end{equation}
Haze produces the vertical gradient with $\beta=1$ when eddy diffusion dominates over the settling.
Thus, strong eddy diffusion acts to steepen the spectral slope.
When the settling is dominant, the gradient depends on how particle sizes and densities vary with altitude.
In the next section, we numerically investigate the haze-produced spectral slopes using a microphysical model.

}

\section{Numerical Investigations of Haze-Produced Spectral Slopes}\label{sec:result}
\subsection{Method}
%%%%%%%%%%%%%%%%%%%%%%%%%%%%%%%%%%%%
\begin{figure*}[t]
\centering
\includegraphics[clip,width=\hsize]{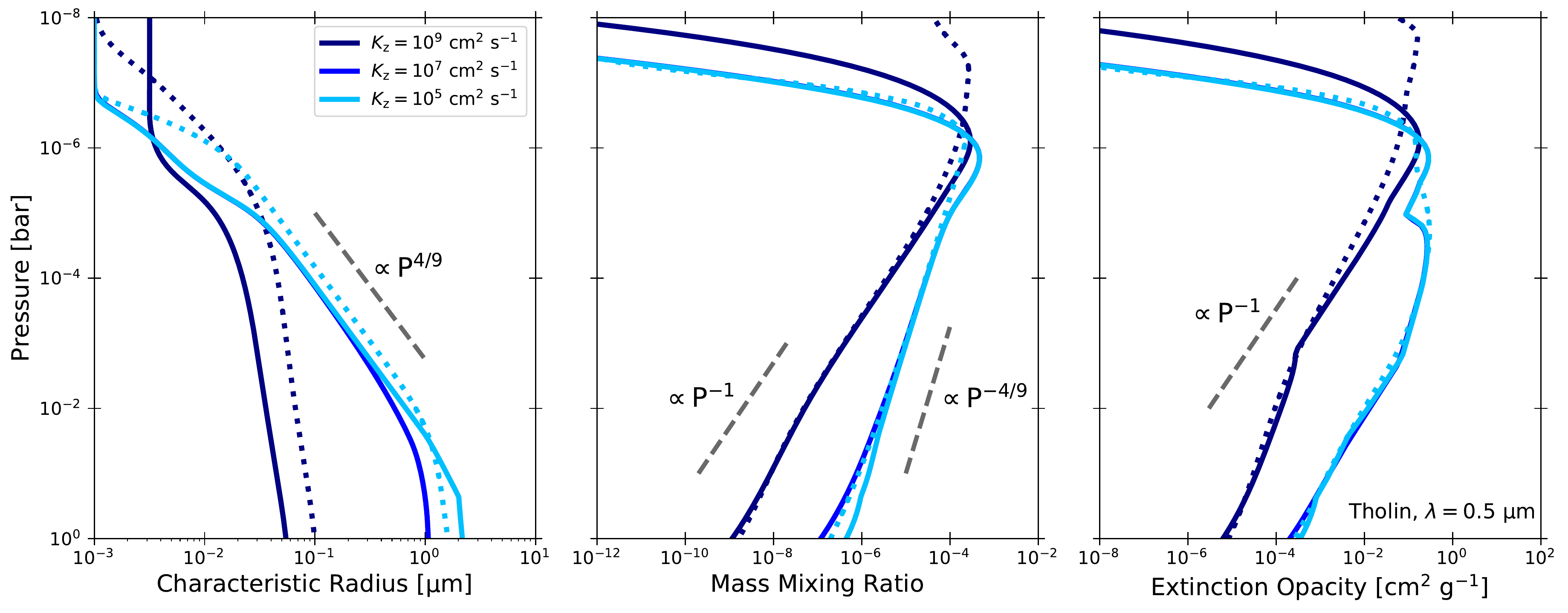}
\caption{
Vertical haze distributions for different eddy diffusion coefficient.
{ From left to right, each column shows the vertical distributions of the characteristic {particle} size, the haze mass mixing ratio, and extinction opacity for the tholin optical constants at $\lambda=0.5~{\rm \mu m}$, respectively.}
%The left and right columns show the vertical distributions of the characteristic {particle} size and the haze mass mixing ratio, respectively.
Different colored lines show the distributions for different $K_{\rm z}$ with haze mass flux $F={10}^{-12}~{\rm g~{cm}^{-2}~s^{-1}}$. %in a top row, and for different $F$ for fixed eddy diffusion coefficient $K_{\rm z}={10}^{7}~{\rm {cm}^2~s^{-1}}$ in a bottom row.
Dotted lines denote the distributions of a volume-weighted particle size predicted by a bin scheme for {$K_{\rm z}={10}^{5}$ and ${10}^{9}~{\rm {cm}^2~s^{-1}}$} taken from Figure 17 of \citet[][]{Kawashima&Ikoma19}, where the column-integrated photolysis rate of haze precursors is {$\sim{10}^{-12}~{\rm g~{cm}^2~s^{-1}}$ (see their Table 1).}
%Vertical profiles of haze mass mixing ratio taken from \citet{Kawashima&Ikoma19}. The purple and green lines show the profiles for $K_{\rm z}={10}^{9}$ and ${10}^{7}~{\rm {cm}^2~s^{-1}}$, respectively. The black dotted line denotes the power law dependence of $q_{\rm h}\propto P^{-1}$, i.e., $\beta=1$. 
}\label{fig:vertical}
\end{figure*}
%%%%%%%%%%%%%%%%%%%%%%%%%%%%%%%%%%%%%

We conduct a series of the calculations for haze particle growth and synthetic transmission spectra.
We utilize a two-moment microphysical model of \citet{Ohno&Okuzumi18} that takes into account the eddy diffusion, gravitational settling, and particle growth.
The moment model suffices to examine whether haze can produce SRSs, as the model can capture the basic effects of haze formation on transmission spectra \citep{Kawashima&Ikoma18}. 
%The basic equations are given by
%\begin{equation}\label{eq:transport_nc}
%\frac{\partial n_{\rm haze}}{\partial t} = \frac{\partial}{\partial z}\left[ \rho_{\rm g}K_{\rm z}\frac{\partial }{\partial z}\left( \frac{n_{\rm haze}}{\rho_{\rm g}}\right)+v_{\rm t}n_{\rm haze} \right] - \left| \frac{\partial n_{\rm haze}}{\partial t} \right|_{\rm coll},
%\end{equation}
%\begin{equation}\label{eq:transport_rhoc}
%\frac{\partial \rho_{\rm haze}}{\partial t} = \frac{\partial}{\partial z}\left[ \rho_{\rm g}K_{\rm z}\frac{\partial }{\partial z}\left( \frac{\rho_{\rm haze}}{\rho_{\rm g}} \right)+v_{\rm t}\rho_{\rm haze} \right],
%\end{equation}
We assume spherical particles with constant density of $1~{\rm g~{cm}^{-3}}$ and ignore the condensation of mineral vapors for the sake of simplicity.
The monomer production profile is prescribed by a log-normal profile given by \citep{Ormel&Min19}
\begin{equation}
    \dot{\rho}_{\rm haze}=\rho_{\rm g}g\frac{F}{\sigma P \sqrt{2\pi}}\exp{\left[ -\frac{1}{2\sigma^2}\left(\ln{\frac{P}{P_{\rm *}}} \right)^2\right]},
\end{equation}
where the characteristic height of monomer production $P_{\rm *}$ and the width of the distribution $\sigma$ are set to $P_{\rm *}={10}^{-6}~{\rm bar}$ and { $\sigma=0.5$} to mimic the profile predicted by photochemical models \citep[e.g.,][]{Kawashima&Ikoma19}.
Correspondingly, we include the increase of a particle number density as $\dot{n}_{\rm haze}=3\dot{\rho}_{\rm haze}/4\pi a_{\rm 0}^3\rho_{\rm p}$, where $a_{\rm 0}$ is the monomer radius and assumed to be $a_{\rm 0}=1~{\rm nm}$.
%The monomer size less affects results for compact particles \citep[e.g.,][]{Adams+19}, and thus we assume $r_{\rm 0}=1~{\rm nm}$. 
The pressure-temperature structure for a solar composition atmosphere is constructed by an analytical model of \citet{Guillot10} {using the same parameters adopted} in \citet{Kawashima&Ikoma19} for their case of irradiation temperature of $790~{\rm K}$ \footnote{The temperature is a product of $\sqrt{2}$ and equilibrium temperature, which characterizes irradiation intensity \citep[see e.g.,][]{Guillot10}. }.

%Using the calculated haze vertical profiles, 
We compute synthetic transmission spectra of hazy atmospheres using a model of \citet{Ohno+19} {assuming the planetary mass of GJ~1214b \citep[$6.26M_{\rm Earth}$,][]{Anglada-Escud+13} and the reference radius of $R_{\rm 0}=2.35R_{\rm Earth}$ at $P=10~{\rm bar}$}.
We introduce a metric quantifying steepness of the spectral slopes defined as \citep{Pinhas&Madhusudhan17}
\begin{equation}
    \mathcal{S}\equiv \frac{1}{H}\frac{dR_{\rm p}}{d\ln{\lambda}}.
\end{equation}
We use the $U$ band ($\lambda=365~{\rm nm}$, FWHM of $66~{\rm nm}$) and $V$ band ($\lambda=551~{\rm nm}$, FWHM of $88~{\rm nm}$) \citep{1998gaas.book.....B} to calculate $\mathcal{S}$, as similar to \citet{Pinhas&Madhusudhan17}.
%The SRS can be defined as the slope with $\mathcal{S}<-4$.
The haze opacity is calculated by the BHMIE \citep{Bohren&Huffman83} assuming spherical particles.
The refractive index has been unknown for {exoplanetary haze}.
We test the two representative refractive index; a Titan haze analog \citep[tholin,][]{Khare+84} and a complex refractory hydrocarbon (soot) compiled by \citet{Lavvas&Koskinen17}.
%For instance, the soot yields more flat-like spectra than those for the tholin \citep[see e.g.,][]{Adams+19}. 

%We vary the downward mass flux $F$ as a free parameter.
%For the sake of simplicity, we neglect the porosity evolution 
%vertical profile depends on how the particle size $r$ varies with altitude.
%According to haze microphysical models, the size decreases with increasing altitude \citep{Lavvas&Koskinen17,Kawashima&Ikoma18,Kawashima&Ikoma19}. 
%Thus, the mixing ratio is expected to increase with increasing altitude.

%%%%%%%%%%%%%%%%%%%%%%%%%%%%%%%%%%%
%\section{Results}\label{sec:result}
\subsection{Haze Vertical Profiles}\label{sec:vertical}
%%%%%%%%%%%%%%%%%%%%%%%%%%%%%%%%%%%
\begin{figure*}[t]
\centering
\includegraphics[clip,width=0.62\hsize]{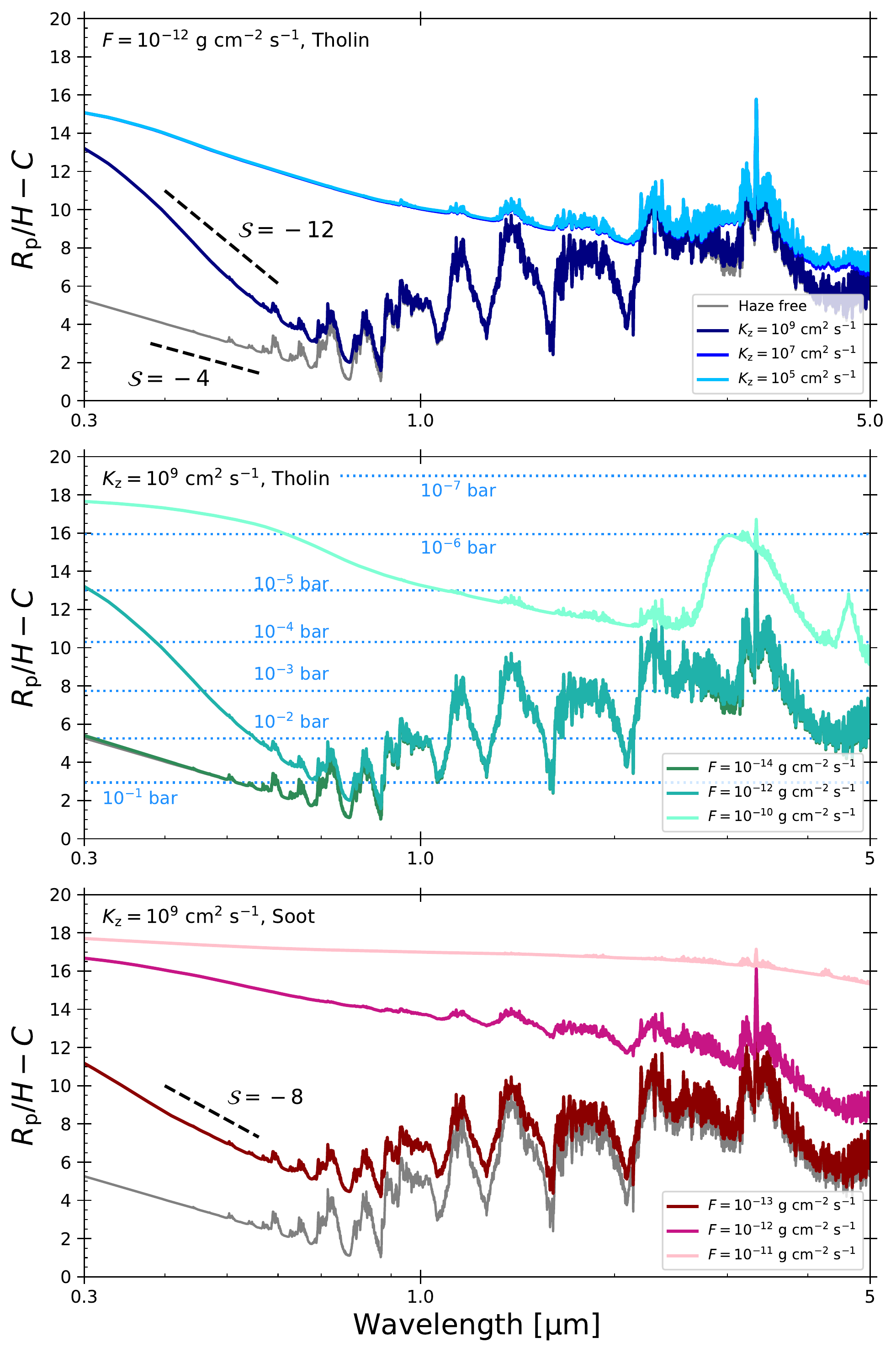}
\caption{
Synthetic transmission spectra {of hazy atmospheres}. The vertical axis is the planetary radius normalized by the scale height of $H=180~{\rm km}${, the value for $P\sim{10}^{-3}~{\rm bar}$,} with an offset. The top panel shows the spectra for different $K_{\rm z}$ with $F={10}^{-12}~{\rm g~{cm}^{-2}~s^{-1}}$. The middle and bottom panels show the spectra for different $F$ with $K_{\rm z}={10}^{9}~{\rm {cm}^{-2}~s^{-1}}$. The tholin refractive index is assumed for the top and middle panels, while the soot refractive index is assumed for the bottom panel. Horizontal dotted lines in the middle panel denote the $R_{\rm p}/H-C$ corresponding to the pressure levels from ${10}^{-1}$ to ${10}^{-7}~{\rm bar}$.
}\label{fig:spectrum}
\end{figure*}
%%%%%%%%%%%%%%%%%%%%%%%%%%%%%%%%%%%
Haze vertical distributions substantially vary with altitude, as suggested by previous studies \citep[e.g.,][]{Lavvas&Koskinen17,Kawashima&Ikoma18,Kawashima&Ikoma19,Kawashima+19,Adams+19,Lavvas+19,Gao&Zhang20}.
Figure \ref{fig:vertical} shows the vertical distributions of haze characteristic size and mass mixing ratio for $F={10}^{-12}~{\rm g~{cm}^{-2}~s^{-1}}$ and different $K_{\rm z}$.
We have confirmed that our two-moment model well reproduces the distributions simulated by the bin scheme (dotted lines) taken from \citet{Kawashima&Ikoma19}.
%, especially for vertical mass distributions. 
In principle, the particle size increases with decreasing the altitude because of collision{al} growth.
The higher eddy diffusion coefficient is, the smaller particle size is.
This is because efficient vertical mixing transports the particles downward before they grow into large sizes \citep{Kawashima&Ikoma19}.
The high eddy diffusion coefficient also produces {a} steep vertical gradient in the mass mixing ratio, as seen in the case of $K_{\rm z}={10}^9~{\rm {cm}^2~s^{-1}}$.
{This results in the steep vertical opacity gradient, as predicted in Section \ref{sec:method}.
}
%The gradient is nearly proportional to $P^{-1}$, in agreement with the analytical prediction {of} Section \ref{sec:method}.

%%%%%%%%%%%%%%%%%%%%%%%%%%%%%%%%%%%
\begin{figure*}[t]
\includegraphics[clip,width=\hsize]{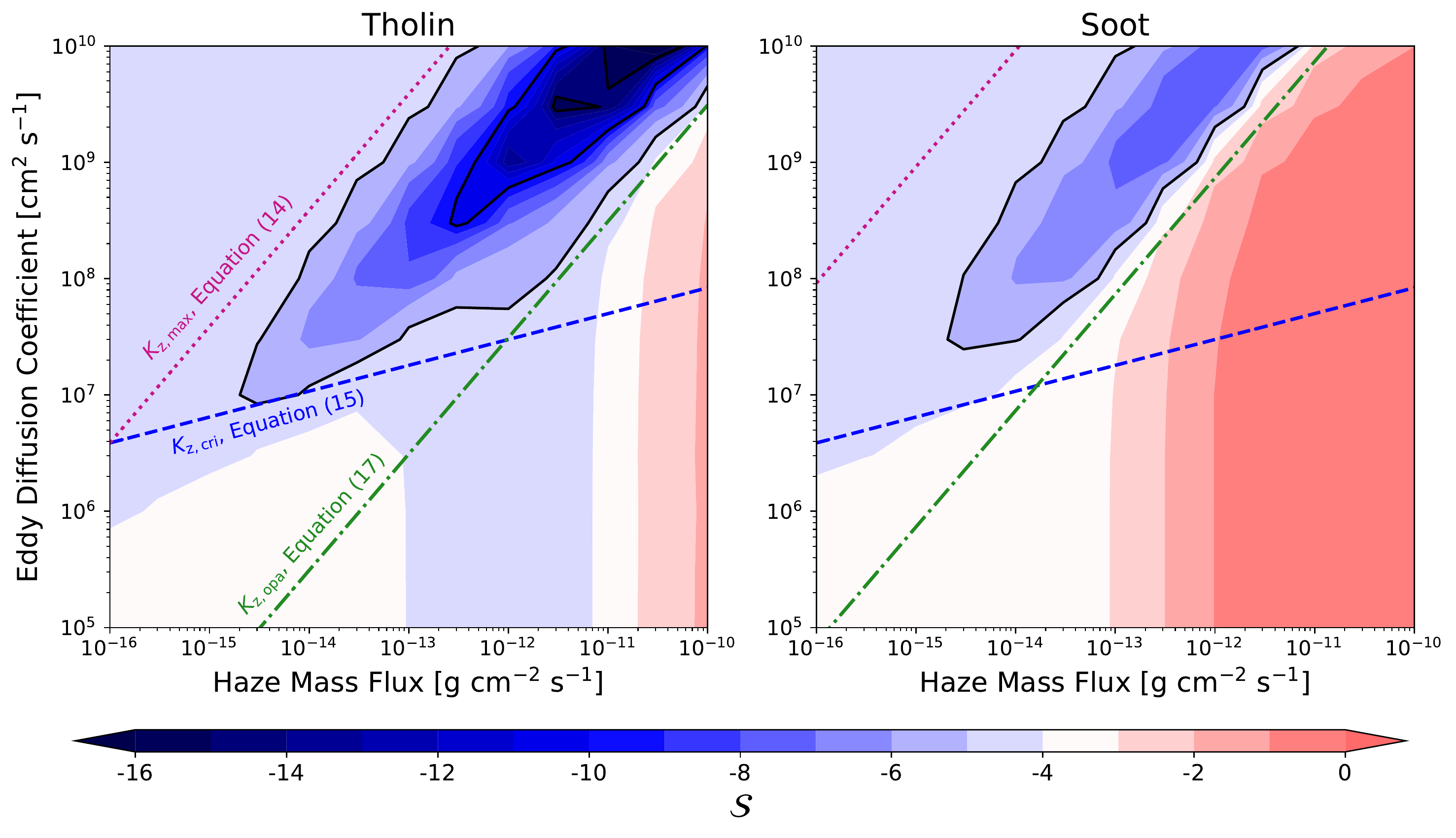}
\caption{
Gradient of the spectral slopes ($\mathcal{S}$, colorscale) as a function of eddy diffusion coefficient and haze mass flux. 
The left and right panels show the results for the tholin and soot haze, respectively. 
The black lines denote the contours of $\mathcal{S}=-5$, $-10$, and $-15$.
The red dotted, blue dashed, and green broken lines denote Equations \eqref{eq:Kz_max}, \eqref{eq:Kz_min}, and \eqref{eq:F_th}, respectively, {for} $P=1~{\rm mbar}$ and {$\lambda=0.55~{\rm \mu m}$. Here, we adopt the Rayleigh scattering cross section of a H$_2$ molecule, $\sigma_{\rm gas}=2.52\times{10}^{-28} ~{\rm {cm}^2}~(\lambda/0.75~{\rm \mu m})^{-4}$ \citep{Levavelier+08}, to evaluate the gas opacity.} %\kappa_{\rm gas}=2.6\times{10}^{-4} ~{\rm {cm}^2~g^{-1}}$ for H$_2$ Rayleigh scattering at $\lambda=0.55~{\rm \mu m}$ \citep{Levavelier+08}.
}\label{fig:diagram}
\end{figure*}

\begin{figure*}[t]
\includegraphics[clip,width=\hsize]{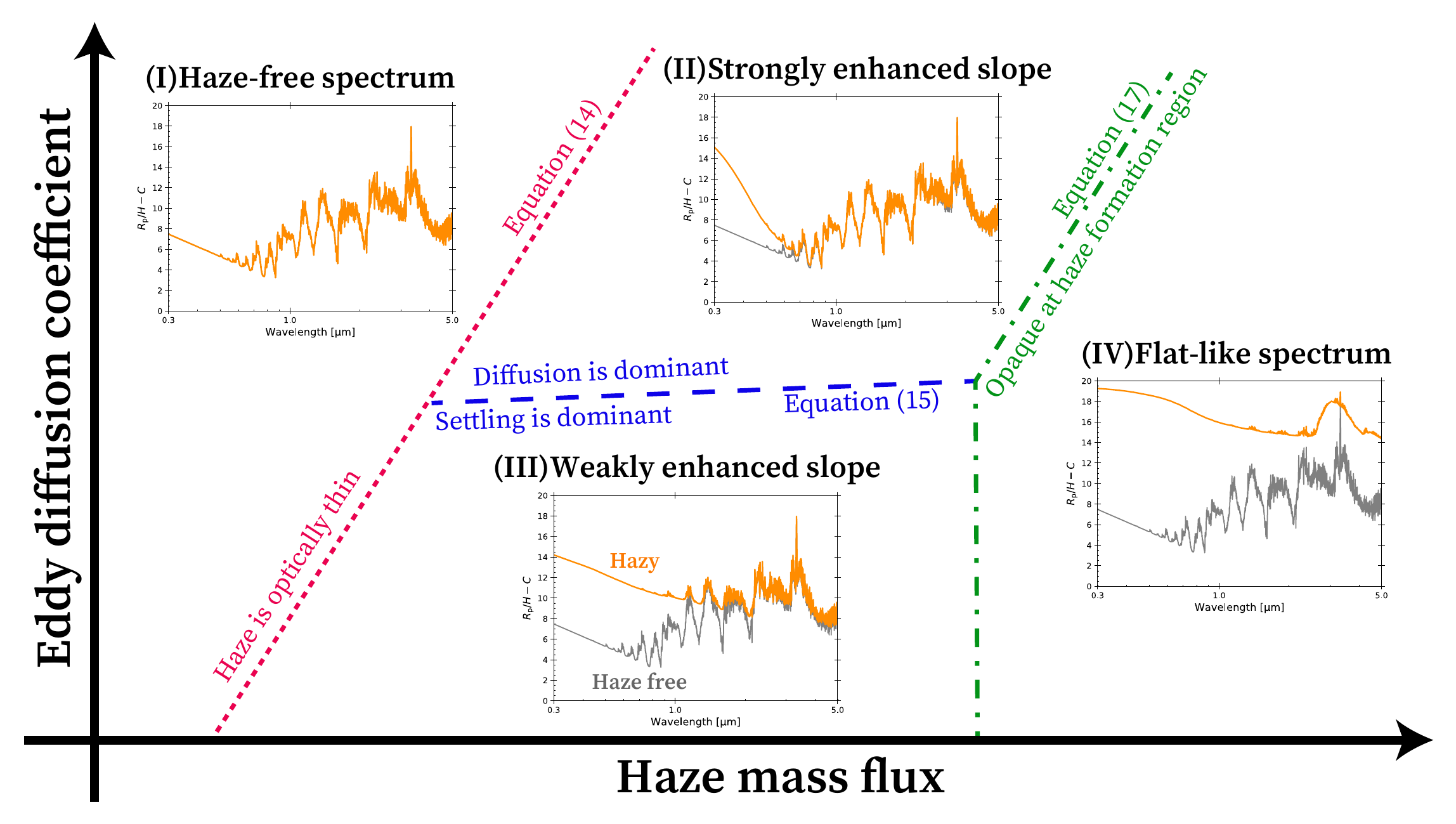}
\caption{Transmission spectrum regimes {in terms} of haze mass flux and eddy diffusion coefficient (see Section \ref{sec:diagram}). Each panel exhibits the typical shape of the transmission spectrum for hazy atmospheres (orange lines) compared to the haze-free spectrum (gray lines).
}\label{fig:regime}
\end{figure*}
%%%%%%%%%%%%%%%%%%%%%%%%%%%%%%%%%%%

The vertical { opacity} gradient also appears when the settling dominates over the eddy diffusion, as seen in the cases of $K_{\rm z}={10}^{7}$ and ${10}^5~{\rm {cm}^2~s^{-1}}$.
This is because the particle size is larger in the deeper atmosphere due to collisional growth, { leading to yield vertical gradient in the mass mixing ratio.} %{and thus the particle settles faster {as compared to the constant size case}}. 
The vertical distributions can be further understood from a timescale argument.
The particle can grow until the settling timescale, $\tau_{\rm settl}=H/v_{\rm t}$, becomes shorter than collision{al} timescale.
For particles smaller than gas mean free path, the collision timescale is approximated by \citep[e.g.,][]{Rossow78}
\begin{equation}\label{eq:tau_coll}
    \tau_{\rm coll}\approx \frac{1}{4}\sqrt{\frac{\rho_{\rm p}}{3ak_{\rm B}T}}\frac{m_{\rm haze}}{\rho_{\rm haze}}=\frac{1}{4}\sqrt{\frac{\rho_{\rm p}}{3ak_{\rm B}T}}\frac{m_{\rm haze}v_{\rm t}}{F},
\end{equation}
where we {have invoked} the mass conservation $\rho_{\rm haze}v_{\rm t}=F$.
%The settling timescale is also estimated as
%\begin{equation}\label{eq:tau_settl}
%    \tau_{\rm sett}=\frac{H}{v_{\rm t}}.
%\end{equation}
Solving $\tau_{\rm coll}=\tau_{\rm sett}$ with Equation \eqref{eq:vt_epstein}, the size is estimated as
\begin{eqnarray}\label{eq:r_coag}
    a&=&\left(\frac{24FP^2}{\pi^2 \rho_{\rm p}^3g^3}\sqrt{\frac{3k_{\rm B}T}{\rho_{\rm p}}} \right)^{2/9}
    %\nonumber
    %&\sim& 0.2~{\rm \mu m}~\left( \frac{F}{{10}^{-12}~{\rm g~{cm}^{-2}~s^{-1}}}\right)^{2/9}\left( \frac{P}{1~{\rm mbar}}\right)^{4/9}\left( \frac{g}{10~{\rm m~s^{-2}}}\right)^{-2/3}%\left( \frac{\rho_{\rm p}}{1~{\rm g~{cm}^{-3}}}\right)^{-7/9},
\end{eqnarray}
%where we have substituted $T=500~{\rm K}$ and $\rho_{\rm p}=1~{\rm g~{cm}^{-3}}$ for an equation in the second row.
Equation \eqref{eq:r_coag} indicates that the particle size is proportional to $P^{4/9}$, which is indeed seen in Figure \ref{fig:vertical}.
Therefore, the mass mixing ratio is proportional to $P^{-4/9}$ for small $K_{\rm z}$ regimes {(see Eqs.~(\ref{eq:vt}) and (\ref{eq:mmr}))}, resulting in the opacity higher at the higher altitude.
{In summary}, the { haze opacity} is higher at higher altitude for all $K_{\rm z}$.
%{ This results in haze opacity that is higher at higher altitude (the right panel of Figure \ref{fig:vertical}), which can enhance spectral slopes.}

%These vertical mass gradients {can be} responsible for producing the SRSs (Section \ref{eq:derivation_slope}).

%A vertical distribution of the mass mixing ratio is also affected by eddy diffusion coefficient.
%Haze mass mixing ratio substantially varies with altitude.
%For $K_{\rm z}={10}^{9}~{\rm {cm}^2~s^{-1}}$, the mass mixing ratio is nearly proportional to the pressure, as predicted from Equation \eqref{eq:mixing_ratio}.
%Since $\rho_{\rm haze}/\rho_{\rm g}\propto F/\rho_{\rm g}v_{\rm t}\propto r^{-1}$, the mass mixing ratio is proportional to $P^{-4/9}$ (see Equation \ref{eq:r_coag}) for small $K_{\rm z}$. 

%Overall trends of the calculated vertical profiles are in agreement with previous studies \citep{Lavvas&Koskinen17,Kawashima&Ikoma18,Kawashima&Ikoma19,Lavvas+19}.
%In particular, we have confirmed that our two-moment model reproduces the distributions simulated by the bin scheme of \citet{Kawashima&Ikoma19} quite well, especially for vertical mass distributions. 
%The largest discrepancy is found in the characteristic size for $K_{\rm z}={10}^{9}~{\rm {cm}^2~s^{-1}}$, which is a factor of $2$--$3$.
%The discrepancy would less affect our subsequent results of synthetic transmission spectra.
%This is because, for large $K_{\rm z}$, the particle size tends to be so small that the opacity can be approximated by the Rayleigh regime, in which the absorption opacity is independent of a particle size.

\subsection{Transmission Spectra}\label{sec:spectrum}

%Haze indeed produces the steep spectral slope, especially for high $K_{\rm z}$, as predicted in Section \ref{sec:method}.
The haze steepens the spectral slope for high $K_{\rm z}$, as found by \citet{Kawashima&Ikoma19}.
The top panel of Figure \ref{fig:spectrum} shows the synthetic transmission spectra for various $K_{\rm z}$ assuming the tholin optical constants. 
%The spectra calculated by the soot optical constants are also shown in the bottom panel.
For the t{case of} $K_{\rm z}={10}^{9}~{\rm {cm}^2~s^{-1}}$, haze produces the spectral slope characterized by $\mathcal{S}\approx-12$ in the optical wavelength, quite steeper than the canonical Rayleigh slope ($\mathcal{S}=-4$).
This stems from the vertical mass gradient produced by efficient eddy diffusion (Section \ref{sec:vertical}).
The spectra for $K_{\rm z}={10}^{7}$ and ${10}^{5}~{\rm {cm}^2~s^{-1}}$ are nearly superposed each other, as the vertical distributions are nearly the same.
%These trends are in agreement with the spectra of \citet{Kawashima&Ikoma19} calculated for the similar haze mass flux (see their Figure 15).

%Does haze produce the steep spectral slope whenever $K_{\rm z}$ is high?
%The answer is no because the spectrum also depends on the haze mass flux.
The haze steepens the spectral slope only when the mass flux fall{s} into {a} moderate value.
As shown in the middle panel of Figure \ref{fig:spectrum}, the steep slope disappears in both cases of high and low mass flux.
The low mass flux ($F={10}^{-14}~{\rm g~{cm}^{-2}~s^{-1}}$) leads to produce {a spectrum superposed on} a haze-free spectrum because the haze { becomes optically thin as compared to} the Rayleigh scattering opacity of H$_2$.
By contrast, the high mass flux ($F={10}^{-10}~{\rm g~{cm}^{-2}~s^{-1}}$) leads to flatten the spectrum because the haze becomes optically thick near the monomer-formation region ($\sim{10}^{-6}~{\rm bar}$) {up to a relatively long wavelength}.
%Therefore, haze steepens the spectral slope only when the mass flux fall{s} into {a} moderate value.

The spectral slope also depends on the optical constants.
The bottom panel of Figure \ref{fig:spectrum} shows the spectra calculated with the soot optical constants.
The soot haze tends to flatten the spectra owing to the weak wavelength dependence of its absorption opacity.
Although the slope is relatively gentle, the soot haze still produces the SRSs with $\mathcal{S}\approx-8$ for $K_{\rm z}={10}^9~{\rm {cm}^2~s^{-1}}$ and $F={10}^{-11}~{\rm g~{cm}^{-2}~s^{-1}}$. 
%{Thus, the quantitative prediction of the spectral slope relies on the assumed optical constants.
%This highlights the importance of} establishing the reliable optical constants of exoplanetary haze from laboratory studies \citep[e.g.,][]{He+18}.
%Although the slope is relatively gentle, the soot haze still potentially produces the SRSs. 
%This is demonstrated in the case of $K_{\rm z}={10}^9~{\rm {cm}^2~s^{-1}}$ and $F={10}^{-11}~{\rm g~{cm}^{-2}~s^{-1}}$ that yields $\mathcal{S}\approx-8$.
%In next section, we investigate what condition leads to produce the steep slope in more detail.

\subsection{In what conditions haze produces SRSs?}\label{sec:diagram}
%We further elaborate what conditions enable haze to produce the {SRS}.
There is a ``sweet spot'' in the $F$--$K_{\rm z}$ space to produce the SRSs.
Figure \ref{fig:diagram} summarizes the spectral slopes calculated for $U$-$V$ bands as a function of haze mass flux $F$ and eddy diffusion coefficient $K_{\rm z}$.
%We find that there is a ``sweet spot'' in the $F$--$K_{\rm z}$ space to produce the SRSs.
The slopes are relatively flat {(i.e., $\mathcal{S}\sim0$)} for very high $F$, as the haze becomes optically thick near the monomer formation region (Section \ref{sec:spectrum}).
By contrast, low $F$ and high $K_{\rm z}$ tend to yield $\mathcal{S}= -4$, as the haze {becomes} optically thin.
{For} moderate mass flux, say $F\ga{10}^{-14}~{\rm g~{cm}^{-2}~s^{-1}}$, the slopes have $\mathcal{S}\sim -4$ for $K_{\rm z}\la {10}^{7}~{\rm {cm}^2~s^{-1}}$ and $\mathcal{S}< -5$ for $K_{\rm z}\ga {10}^{7}~{\rm {cm}^2~s^{-1}}$.
%The large negative value of $\mathcal{S}$ 
{The steep spectral slope (i.e., small $\mathcal{S}$)} for high $K_{\rm z}$ stems from the steep vertical gradient in the mass mixing ratio (Section \ref{sec:vertical}).
In the parameter space examined here, the most steep slope has $\mathcal{S}\approx-16$ for the tholin haze and $\mathcal{S}\approx-8$ for the soot haze, which is found for $F\sim{10}^{-12}$--${10}^{-11}~{\rm g~{cm}^{-2}~s^{-1}}$ and $K_{\rm z}\sim {10}^{10}~{\rm {cm}^2~s^{-1}}$.

%Based on the findings so far, 
%{Here, we demarcate} 
The transmission spectra of hazy atmospheres can be demarcated into four typical regimes {as presented} in Figure \ref{fig:regime}.
When $K_{\rm z}$ is extremely high, the haze { becomes optically thin as compared to} gas opacity, resulting in a haze-free spectrum (regime I).
Equating Equation \eqref{eq:kappa} and the gas opacity $\kappa_{\rm gas}$ with $v_{\rm t}H/K{\rm z}\ll1$, the threshold $K_{\rm z}$ above which the regime I applies is given by
\begin{equation}\label{eq:Kz_max}
    K_{\rm z,max}=\frac{36\pi gH^2F}{\rho_{\rm p}\kappa_{\rm gas}P}\frac{nk\lambda^{-1}}{(n^2-k^2+2)^2+(2nk)^2}.
\end{equation}
%There are two criteria for producing the steep spectral slope.
{For $K_{\rm z}<K_{\rm z,max}$}, the spectrum is substantially affected by haze.
The spectral slope is significantly enhanced by haze when the eddy diffusion dominates over the settling of haze particles (regime II).
Conversely, the slope is only weakly enhanced if the settling dominates over the eddy diffusion (regime III).
Solving $\tau_{\rm diff}=\tau_{\rm sett}$ with Equations \eqref{eq:vt_epstein} and \eqref{eq:r_coag}, where $\tau_{\rm diff}=H^2/K_{\rm z}$ is the diffusion timescale, the critical $K_{\rm z}$ above which eddy diffusion dominates over the settling is estimated as
\begin{eqnarray}\label{eq:Kz_min}
    K_{\rm z,cri}&=&\frac{\rho_{\rm p}g^2H^2}{P\sqrt{8gH/\pi}}\left(\frac{24FP^2}{\pi^2 \rho_{\rm p}^3g^3}\sqrt{\frac{3k_{\rm B}T}{\rho_{\rm p}}} \right)^{2/9}\\
    \nonumber
    &\sim& 3\times{10}^{7}~{\rm {cm}^2~s^{-1}}~\left( \frac{F}{{10}^{-12}~{\rm g~{cm}^2~s^{-1}}}\right)^{2/9}\left( \frac{H}{{200}~{\rm km}}\right)^{3/2}\\
    \nonumber
    &&\times \left( \frac{g}{{10}~{\rm m~s^{-2}}}\right)^{5/6}\left( \frac{P}{{1}~{\rm mbar}}\right)^{-5/9}\left( \frac{T}{{1000}~{\rm K}}\right)^{1/9}\left( \frac{\rho_{\rm p}}{{1}~{\rm g~{cm}^{-3}}}\right)^{2/9}.
\end{eqnarray}
%where we have assumed $\rho_{\rm p}=1~{\rm g~{cm}^{-3}}$ and $T=1000~{\rm K}$.
The spectrum eventually becomes flat when the mass flux is so high that haze is optically thick at the monomer formation region (regime IV).
Since the vertical mass distribution does not follow a power law near the monomer formation region (see Figure \ref{fig:vertical}), we crudely evaluate the optical depth as
\begin{equation}\label{eq:tau_crude}
    \tau_{\rm s}(P_{\rm *})\sim \frac{P_{\rm *}}{gH}\kappa(P_{\rm *}) \sqrt{2\pi R_{\rm p}H}.
\end{equation}
Inserting Equation \eqref{eq:kappa} into \eqref{eq:tau_crude} and solving $\tau_{\rm s}(P_{\rm *})=1$ with $v_{\rm t}H/K_{\rm z}\ll 1$, we achieve the threshold in terms of $K_{\rm z}$ as
\begin{equation}\label{eq:F_th}
    %F_{\rm th}=\frac{\rho_{\rm p}K_{\rm z}}{36\pi HF\sqrt{2\pi R_{\rm p}H}}\frac{\lambda(n^2+2)^2}{nk}.
    K_{\rm z,opa}=\frac{36\pi HF\sqrt{2\pi R_{\rm p}H}}{\rho_{\rm p}}\frac{nk\lambda^{-1}}{(n^2-k^2+2)^2+(2nk)^2}
\end{equation}
Equation \eqref{eq:F_th} does not apply when the settling dominates over the eddy diffusion, i.e., $K_{\rm z}<K_{\rm z,cri}$.
Since the spectrum is invariant with $K_{\rm z}$ for the settling-dominated regime (see Figure \ref{fig:spectrum}), the threshold for $K_{\rm z}<K_{\rm z,cri}$ is given as an intersection of Equations \eqref{eq:Kz_min} and \eqref{eq:F_th}.
We plot Equations \eqref{eq:Kz_max}, \eqref{eq:Kz_min}, and \eqref{eq:F_th} in Figure \ref{fig:diagram} and find that the regime classification well explains the basic behavior of spectral slopes.
Notably, the SRSs preferentially emerge in the regime II (see Figure \ref{fig:regime}){, where} the eddy diffusion coefficient falls into the sweet spot, namely ${\rm max}(K_{\rm z,cri}, K_{\rm z,opa})<K_{\rm z}<K_{\rm z,max}$.
{Alternatively, the SRSs emerge when the haze mass flux falls into a moderate value for given $K_{\rm z}$; for example, $F\sim{10}^{-13}$--${10}^{-11}~{\rm g~{cm}^{-2}~s^{-1}}$ for $K_{\rm z}={10}^{9}~{\rm {cm}^2~s^{-1}}$ (see Figure \ref{fig:diagram}).
Thus, the SRSs might give a constraint on haze mass flux if the strength of eddy diffusion is well constrained.
%$F\sim{10}^{-14}$--${10}^{-11}~{\rm g~{cm}^{-2}~s^{-1}}$, which is in line with the flux predicted by photochemical calculations \citep{Lavvas&Koskinen17,Kawashima&Ikoma19,Lavvas+19}. 
}
%. This regime is applied when the eddy diffusion coefficient falls into the sweet spot, namely ${\rm max}(K_{\rm z,cri}, K_{\rm z,opa})<K_{\rm z}<K_{\rm z,max}$.

%The steep spectral slope indeed emerges at $K_{\rm z,min}<K_{\rm z}<K_{\rm z,max}$ in Figure \ref{fig:diagram}.
%The remarkable feature is that the minimum $K_{\rm z}$ to produce the steep spectral slope only weakly depends on the haze mass flux (Equation \ref{eq:Kz_min}).

\section{Summary and Discussion}\label{sec:summary}
\begin{figure}[t]
\includegraphics[clip,width=\hsize]{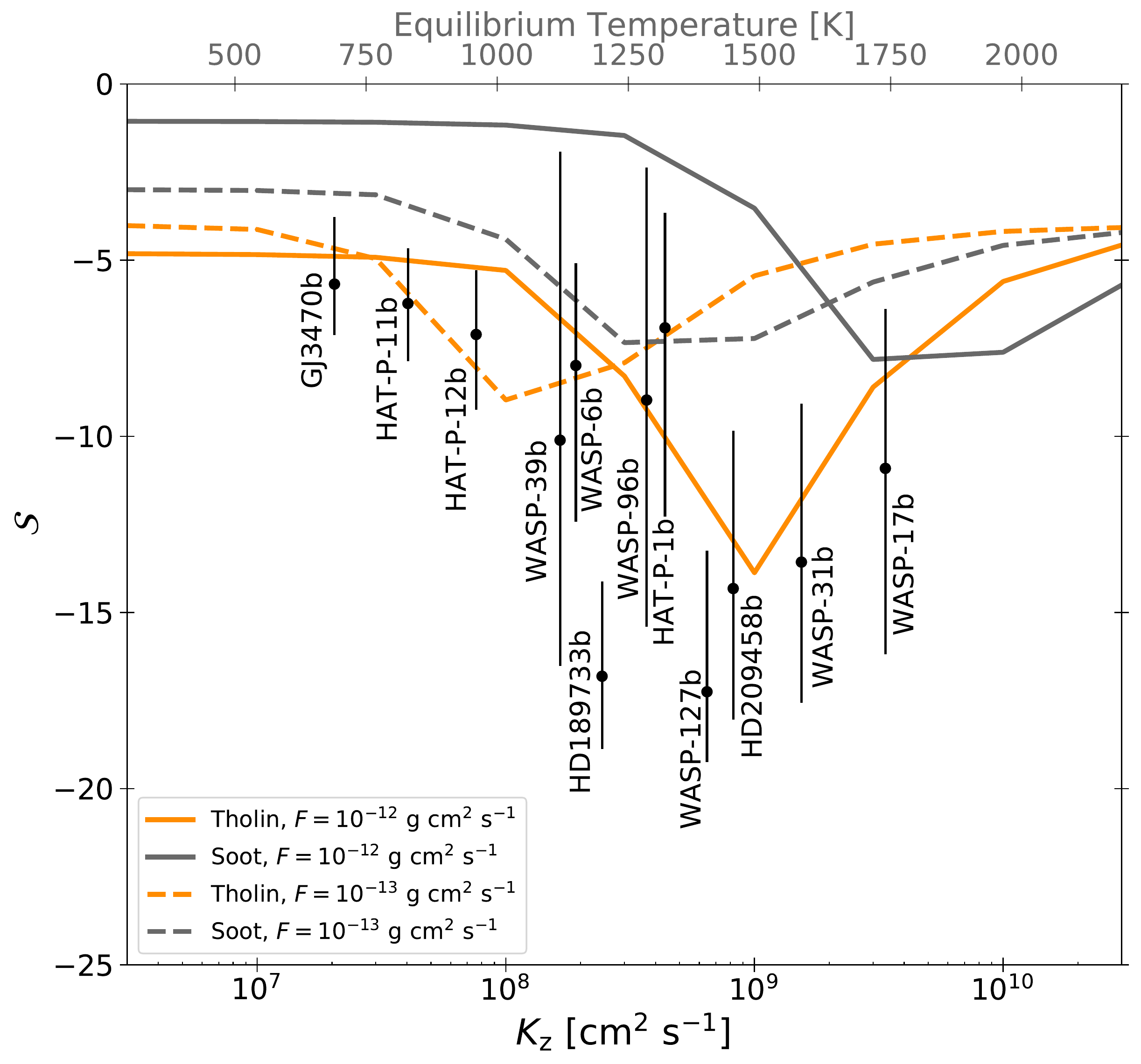}
\caption{Haze-generated spectral slopes as a function of $K_{\rm z}$. The solid and dashed lines show the slopes for $F={10}^{-12}$ and ${10}^{-13}~{\rm g~{cm}^{2}~s^{-1}}$, respectively, which are in line with the predictions from column integrated photolysis rate of hydrocarbons reported by previous studies \citep{Lavvas&Koskinen17,Kawashima&Ikoma19}.
The orange and gray lines exhibit the slopes for tholin and soot haze, respectively. The black dots denote the slopes retrieved by \citet{Welbanks+19} with $1\sigma$ error bars.
}\label{fig:retrieval}
\end{figure}
In this study, we {have} suggested that the super-Rayleigh slopes seen in transmission spectra of some exoplanets can be produced by photochemical haze. %vertical gradient of atmospheric opacity.
We {have} analytically shown that the spectral slope is steepened by the vertical gradient of atmospheric opacity, which is naturally generated by haze (Section \ref{sec:method}). 
%Using a microphysical model, we demonstrated that the opacity gradient is naturally produced by photochemical haze.
We {have} numerically confirmed that the haze can produce the spectral slope several times steeper than the canonical Rayleigh slope, especially when the eddy diffusion coefficient and the haze mass flux fall into the sweet spot (Section \ref{sec:result}).
We {have} also demarcated the transmission spectra of hazy atmospheres into four typical regimes (Figure \ref{fig:regime}).
Our results would help to not only interpret the SRSs but also figure out how haze affects the transmission spectra.

{
One of the possible approaches for testing our idea is to search for the absorption feature of haze itself.
For instance, Titan tholin exhibits absorption features at $3.0$, $4.6$, and $6.3~{\rm {\mu}m}$ \citep[e.g.,][]{Khare+84,Imanaka+04}.
It may be worth investigating whether planets with SRSs show the features at these wavelength.
The actual optical constants of exoplanetary hazes have been unknown.
Therefore, laboratory studies on exoplanetary haze analogs \citep[e.g.,][]{Horst+18,He+18,Moran+20} are important to examine what optical constants are more plausible for exoplanet environments.
The reliable optical constants will also help to quantify the effects of hazes on the spectral slopes.
%It is crucial to establish reliable optical constants of exoplanetary hazes to quantify the impacts on the spectral slopes.
%Laboratory studies for exoplanetary environments will be greatly helpful to this end \citep[e.g.,][]{Horst+18,He+18,Moran+20}.
%If future studies identify the absorption features of exoplanetary haze itself, like tholin's features at $3.0$, $4.6$, and $6.3~{\rm {\mu}m}$ \citep{Khare+84}, it will help to test whether haze causes the SRSs by observing the aerosol features.
}

{
There may be a ``sweet spot'' of planetary equilibrium temperature in which haze preferentially causes the SRSs.
This is because the equilibrium temperature is associated with $K_{\rm z}$ \citep{Komecek+19}.
%specific eddy diffusion coefficient, $K_{\rm z}$, which can be associated to planetary equilibrium temperature.
%through global circulation models \citep[e.g.,][]{Komecek+19}.
%\citet{Welbanks+19} retrieved the spectral-slope gradient $\mathcal{S}$ (their $\gamma$) for exoplanets with various equilibrium temperature, shown in Figure \ref{fig:retrieval}.
%\citet{Komecek+19} found that eddy diffusion coefficient is higher at higher equilibrium temperature.
%This implies that the haze-generated SRSs might preferentially appear at planets with specific equilibrium temperature.
Figure \ref{fig:retrieval} shows the spectral slope $\mathcal{S}$ as a function of $K_{\rm z}$ and corresponding equilibrium temperature, where we have assumed a following relation
\begin{equation}\label{eq:fit_Komacek}
%    \log{_{\rm 10}(K_{\rm z}~{\rm [m^2~s^{-1}])}}=2.110\times{10}^{-3}T_{\rm eq}+1.855.
    \log{_{\rm 10}(K_{\rm z}~{\rm [m^2~s^{-1}])}}=2.110\left(\frac{T_{\rm eq}}{1000~{\rm K}}\right)+1.855.
\end{equation}
This relation is obtained by a linear fit to $K_{\rm z}$ simulated by \citet{Komecek+19} for drag-free atmospheres with $0.01~{\rm \mu m}$ passive tracers at $P=1~{\rm mbar}$ (their Figure 8).
%The haze-generated slope has a local minimum at specific $K_{\rm z}$ because of the transition from regime (III) to regime (I) via regime (II) (see Figure \ref{fig:regime}). 
Haze preferentially produce steep slopes at equilibrium temperature of $\sim1000$--$1500~{\rm K}$ in which $K_{\rm z}$ falls into the regime II. 
Figure \ref{fig:retrieval} also exhibits the slope retrieved by \citet{Welbanks+19} (their $\gamma$).
Interestingly, in the retrieval results, planets with equilibrium temperature of $\sim1200$--$1400~{\rm K}$ tend to exhibit steep spectral slopes, as similar to the haze-generated SRSs.
We do not claim that the result verifies the haze hypotheses since there are many uncertainties, such as stellar contamination \citep[e.g.,][]{McCullough+14} and multi-dimension effects \citep[e.g.,][]{Caldas+19,Macdonald+20}, that should be assessed in future.
Rather, we suggest that measuring the spectral slopes for various equilibrium temperature can help to investigate whether the SRSs are predominantly caused by haze.

%the steepness of spectral slope $\mathcal{S}$ (their $\gamma$) have local minimum at around equilibrium temperature of $\sim1200$--$1400~{\rm K}$ (left panel of Figure \ref{fig:retrieval}), although the uncertainty is large.
%, and the both hotter and cooler planets exhibit relatively gentle slopes. 
%According to a global circulation mode, the eddy diffusion coefficient increases with increasing the equilibrium temperature \citep{Komecek+19}.
%The eddy diffusion coefficient is higher at higher equilibrium temperature because of enhanced atmospheric circulation \citep{Komecek+19}, implying that SRS can preferentially appear at planets with higher equilibrium temperature. 
%The right panel of Figure \ref{fig:retrieval} shows our calculated $\mathcal{S}$ accompanied with the retrieved $\mathcal{S}$ as a function of $K_{\rm z}$, where we convert the equilibrium temperature to $K_{\rm z}$ assuming a relation of

%Equation \eqref{eq:fit_Komacek} is derived by a linear fitting to results of \citet{Komecek+19} for drag-free atmospheres with $0.01~{\rm \mu m}$ passive aerosol tracers at $P=1~{\rm mbar}$ (their Figure 8).
%In Figure \ref{fig:retrieval}, haze-generated SRSs have the local maximum of $\mathcal{S}$ similar to the retrieval results. 

}

%The haze-generated slope may explain the steep spectral slopes of some hot Jupiters.
%{ Photochemical haze also helps to explain the Rayleigh slopes of hot Jupiters, which have been often attributed to mineral clouds.}
Photochemical haze also complements the model of mineral clouds.
Current cloud microphysical models do not predict the optical spectral slope as steep as the Rayleigh slope \citep[][]{Gao&Benneke18,Lines+18b,Lee+19,Powell+19,Ohno+19}, except for \citet{Ormel&Min19} who showed some cases that succeeded in producing the {steep} spectral slopes.
Although it has been believed that haze formation is inefficient in hot exoplanets where CH$_4$ is oxidized to CO \citep{Zahnle+09b}, recent laboratory studies suggest that CO also act as haze precursors \citep{Horst+18,He+19}. 
Thus, haze may still be responsible for hot exoplanets that often show spectral slopes in their transmission spectra.
\acknowledgments 
{ We thank Luis Welbanks for sharing retrieval results of \citet{Welbanks+19}.
We are also grateful to the anonymous referee for constructive comments that greatly improved the quality of this paper.}
K.O. is supported by JSPS KAKENHI Grant Nos. JP18J14557 and JP19K03926.
Y.K. is supported by the European Union’s Horizon 2020 Research and Innovation Programme under Grant Agreement 776403.

\appendix
\section{Derivation of Analytical Haze Opacity}\label{appendix1}
%The vertical distribution of haze mass density $\rho_{\rm haze}$ follows a 1D transport equation given by
%\begin{equation}
%    \frac{\partial \rho_{\rm haze}}{\partial t}=\frac{\partial}{\partial z}\left[\rho_{\rm g}K_{\rm z}\frac{\partial}{\partial z} \left( \frac{\rho_{\rm haze}}{\rho_{\rm g}}\right) +v_{\rm t}\rho_{\rm haze}\right]
%    +\dot{\rho}_{\rm haze},
%\end{equation}
%where $K_{\rm z}$ is the eddy diffusion coefficient, $v_{\rm t}$ is the terminal velocity of haze particles, and $\dot{\rho}_{\rm haze}$ is the mass production rate of haze monomers.
%Here, we analytically demonstrate that photochemical haze {can} produce such gradient. 
In this appendix, we derive the vertical distribution of atmospheric opacity including haze.
The steady vertical distribution of haze mass density $\rho_{\rm haze}$ is determined by the mass conservation, which reads
\begin{equation}\label{eq:steady}
    \frac{K_{\rm z}}{gH^2}P^2\frac{\partial}{\partial P}\left( \frac{\rho_{\rm haze}}{\rho_{\rm g}}\right) -v_{\rm t}\rho_{\rm haze}=-F,
\end{equation}
%where $K_{\rm z}$ is the eddy diffusion coefficient, $g$ is the surface gravity, $v_{\rm t}$ is the terminal velocity of haze particles, and $F$ is the downward mass flux.
{where we have used the hydrostatic equilibrium, the ideal gas law, and the definition of $H$.}
In the upper atmospheres where haze particles are much smaller than gas mean free path, the terminal velocity can be approximated by Equation \eqref{eq:vt_epstein}.
%\begin{equation}\label{eq:vt_epstein_append}
%    v_{\rm t}\approx \frac{\rho_{\rm p}g^2H}{P\sqrt{8k_{\rm B}T/\pi m_{\rm g}}}a.
%    \label{eq:vt}
%\end{equation}
%where $k_{\rm B}$ is the Boltzmann constant, $T$ is the temperature, $m_{\rm g}$ is the mean mass of atmospheric gas particles, $\rho_{\rm p}$ is the internal density of a haze particle, $a$ is the particle radius.
{For} constant $K_{\rm z}$, $\rho_{\rm p}$, and $a$, Equation \eqref{eq:steady} is solved as
\begin{equation}\label{eq:mixing_ratio}
    \frac{\rho_{\rm haze}}{\rho_{\rm g}}=\frac{gHF}{Pv_{\rm t}}\left[1-\exp{\left(-\frac{v_{\rm t}H}{K_{\rm z}} \right)}\right],
    \label{eq:mmr}
\end{equation}
where we {have} set the boundary condition of $\rho_{\rm haze}=0$ at $P=\infty$.
The extinction cross sections of haze particles may be approximated by absorption cross section, especially for particles much smaller than relevant wavelength. 
For such tiny particles, the absorption cross section is approximated by \citep{Bohren&Huffman83,Kataoka+14}
{\begin{equation}\label{eq:sigma_abs}
    %\sigma_{\rm abs} \approx \pi r^2\frac{24nk}{(n^2+2)^2}\frac{2\pi r}{\lambda},
    \sigma_{\rm abs} \approx \pi a^2\frac{24nk}{(n^2-k^2+2)^2+(2nk)^2}\frac{2\pi a}{\lambda}.
\end{equation}}
%where $n$ and $k$ are real and imaginary parts of complex refractive index.
Combining Equations \eqref{eq:mixing_ratio} and \eqref{eq:sigma_abs}, we finally achieve the opacity (Equation \ref{eq:kappa}) as
\begin{eqnarray}\label{eq:kappa_append}
    \kappa &=& \frac{3\sigma_{\rm abs}}{4\pi a^3\rho_{\rm p}}\frac{\rho_{\rm haze}}{\rho_{\rm g}}\\
    \nonumber
    &=&\frac{36\pi gHF}{\rho_{\rm p}} {\underbrace{\frac{1}{Pv_{\rm t}}\left[1-\exp{\left(-\frac{v_{\rm t}H}{K_{\rm z}} \right)}\right]}_{\text{Pressure dependence}}}\underbrace{\frac{nk\lambda^{-1}}{(n^2-k^2+2)^2+(2nk)^2}}_{\text{Wavelength dependence}},
\end{eqnarray}
Equation \eqref{eq:kappa_append} demonstrates that, for tiny absorbing haze, the pressure dependence is originated from the vertical mass gradient, while the wavelength dependence is from haze optical constants.

\end{document}